# Using Table Valued Functions in SQL Server 2005 To Implement a Spatial Data Library

Jim Gray
*Microsoft Research*
Alex Szalay, Gyorgy Fekete,
*The Johns Hopkins University*

# Using Table Valued Functions in SQL Server 2005
# To Implement a Spatial Data Library

Jim Gray, Microsoft (contact author)
Alex Szalay, Johns Hopkins University
Gyorgy Fekete, Johns Hopkins University
August 2005

## Abstract

This article explains how to add spatial search functions (point-near-point and point in polygon) to Microsoft SQL Server™ 2005 using C# and table-valued functions. It is possible to use this library to add spatial search to your application without writing any special code. The library implements the public-domain C# Hierarchical Triangular Mesh (HTM) algorithms from Johns Hopkins University. That C# library is connected to SQL Server 2005 via a set of scalar-valued and table-valued functions. These functions act as a spatial index.

## Resources
The article is illustrated by examples that can be downloaded from http://msdn.microsoft.com/sql/2005/. The sample package includes:
- An 11 MB sample spatial database of United States cities and river-flow gauges.
- The sample queries from the `sql\testScript.sql` article.
- A Visual Studio 2005 project, `\htm.sln`, with all the SQL and C# code.
- A paper, `doc\Table_Valued_Functions.doc`.
- An article, `doc\HtmCsharp.doc`, that provides a manual page for each routine.
- An article, `doc\HTM.doc`, that explains the Hierarchical Triangular Mesh algorithms in detail.
- An article, `doc\There_Goes_the_Neighborhood.doc`, which explains how the HTM algorithms are used in Astronomy. This article also explains two other approaches: zones for batch-oriented point-to-point and point-area comparisons, and regions for doing Boolean algebra on areas. Public domain implementations of those approaches implemented for SQL Server are used in the SkyServer, a popular Astronomy website for the Sloan Digital Sky Survey (http://SkyServer.SDSS.org/ and by several other astronomy data servers.

# Table of Contents

# Introduction

Spatial data searches are common in both commercial and scientific applications. We developed a spatial search system in conjunction with our effort to build the SkyServer (http://skyserver.sdss.org/) for the astronomy community. The SkyServer is a multi-terabyte database that catalogs about 300 million celestial objects. Many of the questions astronomers want to ask of it involve spatial searches. Typical queries include, "What is near this point?" "What objects are inside this area?" and "What areas overlap this area?"

For this article, we have added the latitude/longitude (lat/lon) terrestrial sphere (the earth) grid to the astronomer's right ascension/declination (ra/dec) celestial sphere (the sky) grid. The two grids have a lot in common, but the correspondence is not exact; the traditional order lat-lon corresponds to dec-ra. This order reversal forces us to be explicit about the coordinate system. We call the Greenwich-Meridian-Equatorial terrestrial coordinate system the LatLon coordinate system. The library supports three coordinate systems:
- Greenwich Latitude-Longitude, called LatLon
- Astronomical right-ascension–declination called J2000
- Cartesian (*x, y, z*) called Cartesian.

Astronomers use arc minutes as their standard distance metric. A nautical mile is an arc minute, so the distance translation is very natural. Many other concepts are quite similar. To demonstrate these, this article will show you how to use the spatial library to build a spatial index on two USGS datasets: US cities, and US stream-flow gauges. Using these indexes and some spatial functions, the article provides examples of how to search for cities near a point, how to find stream gauges near a city, and how to find stream gauges or cities within a state (polygonal area).

We believe this approach is generic. The spatial data spine schema and spatial data functions can be added to almost any application to allow spatial queries. The ideas also apply to other multi-dimensional indexing schemes. For example, the techniques would work for searching color space or any other low-dimension metric space.

# Table Valued Functions: The Key Idea

The key concept of relational algebra is that every relational operator consumes one or more relations and produces an output relation. SQL is syntactic sugar for this idea, allowing you to define relations (data definition language) and to manipulate relations with a select-insert-update-delete syntax.

Defining your own scalar functions lets you make some extensions to the relational database – you can send mail messages, you can execute command scripts, and you can compute non-standard scalars and aggregate values such as `tax()` or `median()`.

However, if you can create tables, then you can become part of the relational engine – both a producer and consumer of relational tables. This was the idea of OLEDB, which allows any data source to produce a data stream. It is also the idea behind the SQL Server 2000 Table Valued Functions.

Implementing table valued functions in Transact-SQL is really easy:
```sql
create function t_sql_tvfPoints()
      returns @points table (x float, y float)
 as begin
      insert @points values(1,2);
      insert @points values(3,4);
      return;
      end
```

This is fine if your function can be done entirely in Transact-SQL. But implementing OLEDB data sources or Table Valued Functions outside of Transact-SQL is a real challenge in SQL Server 2000.

The common language runtime (CLR) integration of SQL Server 2005 makes it easy to create a table-valued function. You create a list, an array, or any `IEnumerable` object (anything you can do `foreach` on), and then you cast it as a table. That's all there is to it.

```csharp
[SqlFunction(     TableDefinition  = "x float, y float" ,
                  FillRowMethodName = "FillPair")]
      public static IEnumerable csPoints( )
      {
            int[,] points = { { 1, 2 }, { 3, 4 } };
            return (IEnumerable) points;
      }
```

You compile this in Visual Studio and click **Deploy**. The table-valued function is installed in the database.

## Using Table-Valued Functions to Add a Spatial Index

There is a lot of confusion about indexes. Indexes are really simple – they are tables with a few special properties:
- SQL Server has only one kind of associative (by value) index – a B-tree. The B-tree can have multi-field keys, but the first field carries most of the selectivity.
- Conceptually, the B-Tree index is a table consisting of the B-Tree key fields, the base table key fields, and any included fields that you want to add to the index.
- B-tree indexes are sorted according to the index key, such as.ZIP code or customer ID, so that lookup or sequential scan by that key is fast.
- Indexes are often smaller than the base table, carrying only the most important attributes, so that looking in the index involves many fewer bytes than examining the whole table. Often, the index is so much smaller that it can fit in main memory, thereby saving even more disk accesses.
- When you think you are doing an index lookup, you are either searching the index alone (a vertical partition of the base table), or you are searching the index, and then joining the qualifying index rows to rows in the base table via the base-table primary key (a bookmark lookup).

The central idea is that the spatial index gives you a small subset of the data. The index tells you where to look and often carries some helpful search information with it (called included columns or covering columns by the experts.) The selectivity of an index tells how big this initial reduction is (the coarse subset of Figure 1). When the subset is located, a careful test examines each member of the subset and discards false positives. That process is indicated by the diamond in Figure 1. A good index has few false positives. We use the Figure 1 metaphor (the coarse subset and the careful test) throughout this article.

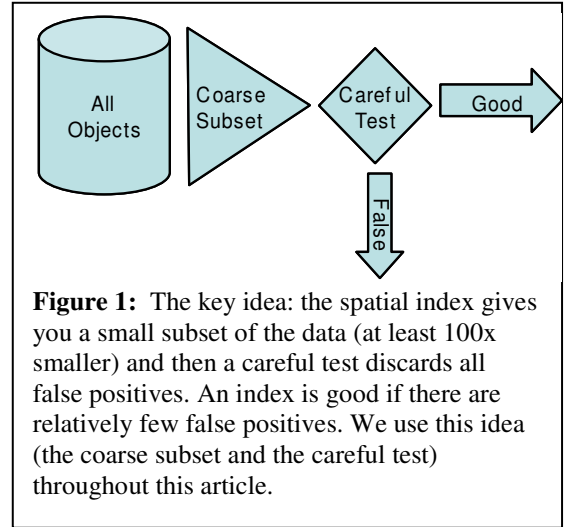

**Figure 1:** The key idea: the spatial index gives you a small subset of the data (at least 100x smaller) and then a careful test discards all false positives. An index is good if there are relatively few false positives. We use this idea (the coarse subset and the careful test) throughout this article.

B-trees and table-valued functions can be combined as follows to let you build your own spatial index that produces coarse subsets:
1. Create a function that generates keys that cluster related data together. For example, if items A and B are related, then the keys for A and B should be nearby in the B-tree key space.
2. Create a table-valued function that, given a description of the subset of interest, returns a list of key ranges (a "cover") containing all the pertinent values.

You cannot always get every key to be near all its relatives because keys are sorted in one dimension and relatives are near in two-dimensional space or higher. However, you can come close. The ratio of false-positives to correct answers is a measure of how well you are doing.

The standard approach is to find some space filling curve and thread the key space along that curve. Using the standard Mercator map, for example, you can assign everyone in the Northwest to the Northwest key range, and assign everyone in the Southeast to the Southeast key range. Figure 2 shows the $2^{nd}$ order space-filling curve that traverses all these quadrants, assigning keys in sequence. Everyone in the Northwest-Southwest quadrant has the key prefix `nwsw`. If you have an area like the circle shown in Figure 2, you can look in the key range
   `key between 'nwsw' and 'nwse'`
This search space is eight times smaller than the whole table and has about 75 percent false positives (indicated by the area outside the circle but inside the two boxes). This is not a great improvement, but it conveys the idea. A better index would use a finer cell division. With fine enough cells, the converging area could have very few false positives. A detailed review of space-filling curves and space-partitioning trees can be found in the books of Hanan Samet [Samet].

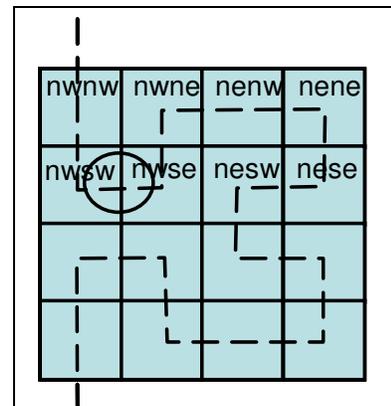

**Figure 2:** The start of a space filling Peano or Hilbert curve (one recursively divides each cell in a systematic way.) The cells are labeled. All points in cell 'nwse' have a key with that prefix so you can find them all in the 'nwse' section of the B-tree, right before 'nwne' and right after 'nwsw'. The circle is an area of interest that overlaps two such cells.

Now we are going to define a space-filling curve – the Hierarchical Triangular Mesh (HTM) that works particularly well on the sphere. The earth is round and the celestial sphere is round, so this spherical system is very convenient for geographers and astronomers. We could do similar things for any metric space. The space-filling curve gives keys that are the basis of the spatial index. Then, when someone has a region of interest, our table valued function will give them a good set of key-ranges to look at (the coarse filter of Figure 1). These key ranges will cover the region with spherical triangles, called trixels, much as the two boxes in Figure 2 cover the circle.  The search function need only look at all the objects in the key ranges of these trixels to see if they qualify (the careful test in Figure 1).

To make this concrete, assume we have a table of Objects
```sql
create table Object (   objID bigint primary key,
                        lat   float, -- latitude
                        lon   float, -- longitude
                        HtmID bigint)-- the HTM key
```
and a distance function `dbo.fDistanceLatLon(lat1, lon1, lat2, lon2)` that gives the distance in nautical miles (arcminutes) between two points. Further assume that the following table-valued function gives us the list of key ranges for HtmID points that are within a certain radius of a lat-lon point.
```sql
define function
    fHtmCoverCircleLatLon(@lat float, @lon float, @radius float)
    returns @TrixelTable table(HtmIdStart bigint, HtmIdEnd bigint)
```

Then the following query finds points within 40 nautical miles of San Francisco (lat,lon) = (37.8,-122.4):
```sql
select O.ObjID, dbo.fDistanceLatLon(O.lat,O.lon, 37.8, -122.4)
from fHtmCoverCircleLatLon(37.8, -122.4, 40) as TrixelTable
 join Object O
   on O.HtmID between TrixelTable.HtmIdStart         -- coarse test
                  and TrixelTable.HtmIdEnd
where dbo.fDistanceLatLon(lat,lon,37.8, -122.4) < 40   -- careful test
```

We now must define the HTM key generation function, the distance function, and the HTM cover function.  That's what we do next using two United States Geological spatial datasets as an example. If you are skeptical that this scales to billions of objects, go to http://skyserver.sdss.org/ and look around the site. That Web site uses this same code to do its spatial lookup on a multi-terabyte astronomy database.

This article is about how you use SQL Table Valued Functions and a space-filling curve like the HTM to build a spatial index. As such, we treat the HTM code itself as a black box documented elsewhere [Szalay], and we focus on how to adapt it to our needs within an SQL application.

## The Datasets
The US Geological Survey gathers and publishes data about the United States. Figure 3 shows the locations of 18,000 USGS-maintained stream gauges that measure river water flows and levels. The USGS also publishes a list of 23,000 place names and their populations.

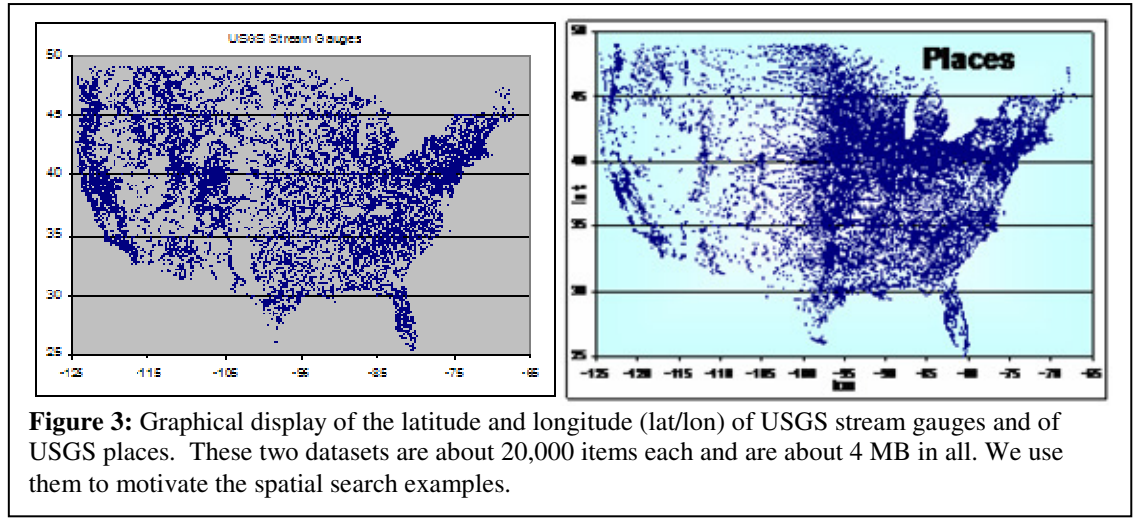

**Figure 3:** Graphical display of the latitude and longitude (lat/lon) of USGS stream gauges and of USGS places. These two datasets are about 20,000 items each and are about 4 MB in all. We use them to motivate the spatial search examples.

## *USGS Populated Places (23,000 cities)*

The USGS published a list of place names and some of their attributes in 1993. There are newer lists at the USGS website but they are fragmented by state, so it is difficult to get a nationwide list. The old list will suffice to demonstrate spatial indicies. The data has the following format:

```
create table Place(
    PlaceName   varchar(100) not null, -- City name
    State       char(2)      not null, -- 2 char state code
    Population  int          not null, -- Number of residents (1990)
    Households  int          not null, -- Number of homes (1990)
    LandArea    int          not null, -- Area in sqare KM
    WaterArea   int          not null, -- water area within land area
    Lat         float        not null, -- latitude in decimal degrees
    Lon         float        not null, -- longitude decimal degrees
    HtmID       bigint       not null primary key --spatial index key
)
```

To speed name lookups, we add a name index, but the data is clustered by the spatial key. Nearby objects are co-located in the clustering B-tree and thus on the same or nearby disk pages.

```
create index Place_Name on Place(PlaceName)
```

All except the `HtmID` data can be downloaded from the USGS Web site. The SQL Server 2005 data import wizard can be used to import the data (we have already done that in the sample database.) The `HtmID` field is computed from the Lat Lon by:

```
update Place set HtmID = dbo.fHtmLatLon(lat, lon)
```

## *USGS Stream Gauges (17,000 instruments)*

The USGS has been maintaining records of river flows since 1854. As of 1 Jan 2000, they had accumulated over 430 thousand years of measurement data. About six thousand active stations were active, and about four thousand were online. The gauges are described in detail at http://waterdata.usgs.gov/nwis/rt. A NOAA site shows the data from a few hundred of the most popular stations in a very convenient way: http://weather.gov/rivers_tab.php.

Our database has just the stations in the continental United States (see Figure 3). There are also stations in Guam, Alaska, Hawaii, Puerto Rico, and the Virgin Islands that are not included in this database. The stream gauge station table is:

```sql
create table Station (
    StationName    varchar(100) not null,      -- USGS Station Name
    State          char(2)      not null,      -- State location
    Lat            float        not null,      -- Latitude in Decimal
    Lon            float        not null,      -- Longitude in Decimal
    DrainageArea   float        not null,      -- Drainage Area (km2)
    FirstYear      int          not null,      -- First Year operation
    YearsRecorded  int          not null,      -- Record years (at Y2k)
    IsActive       bit          not null,      -- Was it active at Y2k?
    IsRealTime     bit          not null,      -- On Internet at Y2K?
    StationNumber  int          not null,      -- USGS Station Number
    HtmID          bigint       not null,      -- HTM spatial key
                                               -- (based on lat/lon)

    primary key(htmID, StationNumber) )
```

As before, the `HtmID` field is computed from the Lat Lon fields by:
```sql
update Station set HtmID = dbo.fHtmLatLon(lat, lon)
```

There are up to 18 stations at one location, so the primary key must include the station number to make it unique. However, the HTM key clusters all the nearby stations together in the B-tree. To speed lookups, we add a station number and a name index:
```sql
create index Station_Name   on Station(StationName)
create index Station_Number on Station(StationNumber)
```

## *The Spatial Index Table*

Now we are ready to create our spatial index.  We could have added the fields to the base tables, but to make the stored procedures work for many different tables, we found it convenient to just mix all the objects together in one spatial index. You could choose (type,HtmID) as the key to segregate the different types of objects; but, we chose (HtmID, key) as the key so that nearby objects of all types (cities and steam gagues) are clustered together.  The spatial index is:

```sql
create table SpatialIndex (
     HtmID    bigint   not null , -- HTM spatial key (based on lat/lon)
     Lat      float    not null , -- Latitude in Decimal
     Lon      float    not null , -- Longitude in Decimal
     x        float    not null , -- Cartesian coordinates,
     y        float    not null , -- derived from lat-lon
     z        float    not null , --,
     Type     char(1)  not null , -- place (P) or gauge (G)
     ObjID    bigint   not null , -- object ID in table
     primary key (HtmID, ObjID) )
```

The Cartesian coordinates will be explained later in this topic. For now, it is enough to say that the function fHtmCenterPoint(HtmID) returns the Cartesian (x,y,z) unit vector for the centerpoint of that HTM triangle. This is the limit point of the HTM, as the center is subdivided to infinitely small trixels.

The SpatialIndex table is populated from the Place and Station tables as follows:
```sql
insert SpatialIndex
select     P.HtmID, Lat, Lon, XYZ.x, XYZ.y, XYZ.z,
           'P' as type, P. HtmID as ObjID
  from   Place P cross apply fHtmLatLonToXyz(P.lat, P.lon)XYZ

insert SpatialIndex
  select S.HtmID, Lat, Lon, XYZ.x, XYZ.y, XYZ.z,
           'S' as type, S.StationNumber as ObjID
  from   Station S cross apply fHtmLatLonToXyz(S.lat, S.lon) XYZ
```

To clean up the database, we execute:
```sql
DBCC INDEXDEFRAG    ( spatial  , Station, 1)
DBCC INDEXDEFRAG    ( spatial  , Station, Station_Name)
DBCC INDEXDEFRAG    ( spatial  , Station, Station_Number)
DBCC INDEXDEFRAG    ( spatial  , Place,   1)
DBCC INDEXDEFRAG    ( spatial  , Place,   Place_Name)
DBCC INDEXDEFRAG    ( spatial  , SpatialIndex, 1)
DBCC SHRINKDATABASE ( spatial  , 1  ) -- 1% spare space
```

## *A Digression: Cartesian Coordinates*

You can skip this if you like. It is not needed to use the library. The HTM code heavily uses a trick to avoid spherical geometry: it moves from the 2D surface of the sphere to 3D. This allows very quick tests for "inside a polygon" and for "nearby a point" queries.

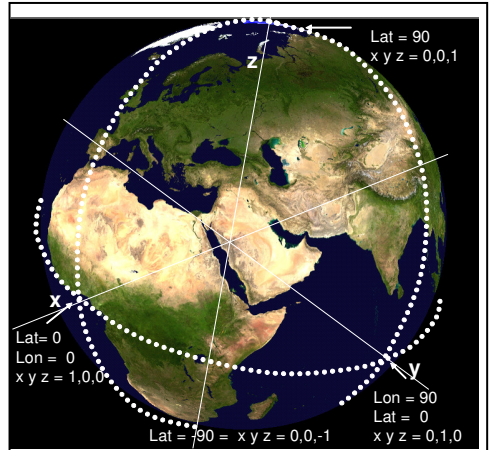

Every lat/lon point on the sphere can be represented by a unit vector in three-dimensional space **v** = *(x,y,z)*. The north and south poles (90° and -90°) are **v** = (0,0,1), and **v** = (0,0,-1) respectively. *Z* represents the axis of rotation, and the *XZ* plane represesnt the Prime (Greenwich) Meridian ,having longitude 0° or longitude 180°. The formal definitions are:

> $x = \cos(lat)\cos(lon)$
> $y = \cos(lat)\sin(lon)$
> $z = \sin(lat)$

**Figure 4:** Cartesian coordinates allow quick tests for point-in-polygon and point-near-point. Each lat/lon point has a corresponding *(x,y,z)* unit vector.

These Cartesian coordiates are used as follows. Given two points on the unit sphere, **p1**=$(x_1,y_1,z_1)$ and **p2** = $(x_2,y_2,z_2)$, then their dot product, **p1•p2** = $x_1 * x_2 + y_1 * y_2 + z_1 * z_2$, is the cosine of the angle between these two points. It is a distance metric.

If we are looking for points within 45 nautical miles (arc minutes) of point **p1**, that is at most 45/60 degrees away from **p1.** The dot product of such points with **p1** will be less than d=cos(radians(45/60)*.* The "is nearby" test becomes { **p2** | **p2•p1** < *d*}, which is a very quick test.

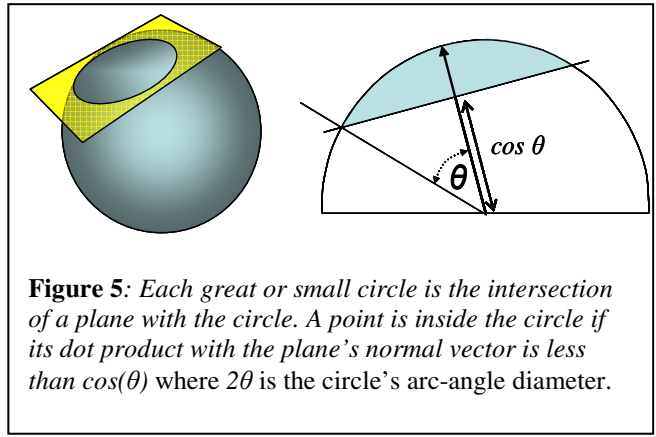

Cartesian coordinates also allow a quick test for point-inside-polygon. All our polygons have great-circle or small-circle edges. Such edges lie along a plane intersecting the sphere. So the edges can be defined by the unit vector, **v**, normal to the plane and by a shift along that vector. For example, the equator is the vector **v** = (0,0,1) and shift zero. Latitude 60° is defined by vector **v** = (0,0,1) with a shift of 0.5, and a 60° circle around Baltimore is defined by vector **v** = (0.179195, -0.752798, 0.633392) with a shift of 0.5. A place, **p2**, is within 60° of Baltimore if **p2•v** < 0.5*.* The same idea lets us decide if a point is inside or outside a HTM triangle by evaluating three such dot products. That is one of the main reasons the HTM code is so efficient and fast.

**Figure 5**: *Each great or small circle is the intersection of a plane with the circle. A point is inside the circle if its dot product with the plane's normal vector is less than cos(θ) where 2θ is the circle's arc-angle diameter.*

We have implemented several helper procedures to convert from LatLon to Cartesian coordiantes:

    `fHtmXyz(HtmID)` returns the *xyz* vector of the centerpoint of an HtmID
    `fHtmLatLonToXyz(lat,lon)` returns an *xyz* vector
    `fHtmXyzToLatLon(x,y,z)` returns a lat,lon vector.

They are used below and documented in the the API spec and Intellisense [Fekete].

The library here defaults to 21-deep HTM keys (the first level divides the sphere into 8 faces and each subsequent level divides the speherical triangle into 4-sub-triangles.) The table below indicates that a 21-deep trixel is fairly small. The code can be modified to go 31-deep deep before the 64-bit representation runs out of bits.

**Table 1:** Each HTM level subdivdes the sphere. For each level, this table shows the area in square degrees, arc minutes, arc seconds, and meters. The Trixel colum shows some charactic sizes: the default 21-deep trixels is about .3 arc second$^2$. The USGS data has about ½ object per 12-deep trixel.

| HTM depth | Area | | | | | objects / trixel | |
|---|---|---|---|---|---|---|---|
| | deg$^2$ | arc min$^2$ | arc sec$^2$ | earth m$^2$ | trixel | SDSS | USGS |
| sphere | 41253 | 148,510,800 | 534,638,880,000 | 5.E+14 | | | |
| 0 | 5157 | 18,563,850 | 66,829,860,000 | 6E+13 | | 3E+8 | |
| 1 | 1289 | 4,640,963 | 16,707,465,000 | 2E+13 | | 8E+7 | |
| 2 | 322 | 1,160,241 | 4,176,866,250 | 4E+12 | | 2E+7 | |
| 3 | 81 | 290,060 | 1,044,216,563 | 1E+12 | | 5E+6 | |
| 4 | 20 | 72,515 | 261,054,141 | 2E+11 | | 1E+6 | 30,000 |
| 5 | 5 | 18,129 | 65,263,535 | 6E+10 | | 3E+5 | 7,500 |
| 6 | 1 | 4,532 | 16,315,884 | 2E+10 | 1 deg$^2$ | 73242 | 1,875 |
| 7 | 3E-1 | 1,133 | 4,078,971 | 4E+9 | | 18311 | 468 |
| 8 | 8E-2 | 283 | 1,019,743 | 1E+9 | | 4578 | 117 |
| 9 | 2E-2 | 71 | 254,936 | 2E+8 | | 1144 | 29 |
| 10 | 5E-3 | 18 | 63,734 | 6E+7 | | 286 | 7 |
| 11 | 1E-3 | 4 | 15,933 | 2E+7 | | 72 | 2 |
| 12 | 3E-4 | 1 | 3,983 | 4E+6 | 1 amin$^2$ | 18 | 0.5 |
| 13 | 8E-5 | 3E-1 | 996 | 943816 | | 4 | 0.1 |
| 14 | 2E-5 | 7E-2 | 249 | 235954 | | 1 | |
| 15 | 5E-6 | 2E-2 | 62 | 58989 | | 0.3 | |
| 16 | 1E-6 | 4E-3 | 16 | 14747 | | . | |
| 17 | 3E-7 | 1E-3 | 4 | 3687 | | | |
| 18 | 8E-8 | 3E-4 | 1 | 922 | | | |
| 19 | 2E-8 | 7E-5 | 2E-1 | 230 | 1 asec$^2$ | | |
| 20 | 5E-9 | 2E-5 | 6E-2 | 58 | 1 km$^2$ | | |
| 21 | 1E-9 | 4E-6 | 2E-2 | 14 | | | |
| 22 | 3E-10 | 1E-6 | 4E-3 | 4 | | | |
| 23 | 7E-11 | 3E-7 | 9E-4 | 1 | 1 m$^2$ | | |
| 24 | 2E-11 | 7E-8 | 2E-4 | 2E-1 | | | |
| 25 | 5E-12 | 2E-8 | 6E-5 | 6E-2 | | | |
| 26 | 1E-12 | 4E-9 | 1E-5 | 1E-2 | | | |

# Typical Queries

Assuming we can get the functions defined, we are ready to do a few queries.

## 1. Find Points near point: find towns near a place.

The most common query is to find all places nearby a certain place or point. Consider the query, "Find all towns within 100 nautical miles of Baltimore Maryland." The HTM triangles covering a 100 nautical mile circle (100 arc minutes from) Baltimore are obtained by

```
select *        -- find a HTM cover 100 NM around Baltimore
from fHtmCoverCircleLatLon(39.3, -76.6, 100)
```

This returns the Trixel Table at right. That is, the `fHtmCoverCircleLatLon()` function returns a set of HTM triangles that "cover" the circle (in this case, a single trixel). The HTM keys of all objects inside the

| The Baltimore circle HTM cover | |
| --- | --- |
| HtmIdStart | HtmIdEnd |
| 14023068221440 | 14027363188735 |

circle are also inside one of these triangles. Now we need to look in all those triangles and discard the false positives (the careful test of Figure 1). We will order the answer set by the distance from Baltimore, so that if we want the closest place, we can just select the `TOP 1 WHERE` distance > 0 (we want to exclude Baltimore itself from being closest).

```
declare @lat float, @lon float
select @lat = lat, @lon = lon
from Place
where Place.PlaceName = 'Baltimore'
  and State = 'MD'
select ObjID, dbo.fDistanceLatLon(@lat,@lon, lat, lon) as distance
from SpatialIndex join fHtmCoverCircleLatLon(@lat, @lon, 100)
   On HtmID between HtmIdStart and HtmIdEnd         -- coarse test
   and type = 'P'
   and dbo.fDistanceLatLon(@lat,@lon, lat, lon) < 100 -- careful test
      order by distance asc
```

The cover join returns 2,928 rows (the coarse test); 1,122 of them are within 100 air miles (the careful test). This gives us 61% false positives – all within 9 milliseconds.

These are such common tasks that there are standard functions for them:

```
fHtmNearbyLatLon(type, lat, lon, radius)
fHtmNearestLatLon(type, lat, lon)
```

so the query above becomes:

```
select ObjID, distance
from fHtmNearestLatLon('P', 39.3, -76.61)
```

## 2. Find places inside a box.

Applications often want to find all the objects inside a square view-port when displaying a square map or window. Colorado is almost exactly square with corner points (41N, -109°3'W) in the NW corner and (37°N-102° 3'E) in the SW corner. The state's center point is (39°N, -105°33'E) so one can cover that square with a circle centered at that point.

```sql
declare @radius float
set @radius = dbo.fDistanceLatLon(41,-109.55,37,-102.05)/2
select *
from Station
where StationNumber in (
    select ObjID
     from fHtmCoverCircleLatLon(39, -105.55, @radius) join SpatialIndex
            on HtmID between HtmIdStart and HtmIdEnd
       and lat between 37 and 41
       and lon between -109.05 and -102.048
       and type = 'S')
OPTION(FORCE ORDER)
```

This example returns 1,030 stream gauges in about 46 milliseconds. Five other Colorado gauges are right on the border that wanders south of 37° by up to 1 nautical mile. These extra 5 stations appear when the southern latitude is adjusted from 37° to 36.98°[1]. The cover circle returns 36 triangles. The join with the SpatialIndex table returns 1,975 gauges. That's 47 percent false positives. The next section shows how to improve on this by using the HTM regions to specify a polygon cover rather than a cover for a circle.

The FORCE ORDER clause is an embarrassment – if missing, the query runs ten times longer because the optimizer does a nested-loops join with the spatial index as the outer table. Perhaps if the tables were larger (millions of rows), the optimizer would pick a different plan, but we cannot count on that. Paradoxically, the optimizer chose the correct plan without any hints for all the queries in the previous section.

---

[1] GIS systems and astronomical applications often want a buffer zone around a region. The HTM code includes support for buffer zones, and they are much used in real applications, Look at reference [Szalay] to see how this is done.

## 3. Find places inside a polygon.

The HTM code lets us specify the area as a circle, a rectangle, a convex hull, or a union of these regions. In particular, the HTM library allows you to specify a region using the following linear syntax:

```
circleSpec  := 'CIRCLE LATLON '       lat lon radius
            |  'CIRCLE J2000 '        ra  dec radius
            |  'CIRCLE [CARTESIAN ]'  x y z   radius
rectSpec    := 'RECT LATLON '         { lat lon }2
            |  'RECT J2000 '          { ra  dec }2
            |  'RECT [CARTESIAN ]'    { x y z   }2
hullSpec    := 'CHULL LATLON '        { lon lat }3+
            |  'CHULL J2000 '         { ra dec  }3+
            |  'CHULL [CARTESIAN ]'   { x y z   }3+
convexSpec  := 'CONVEX ' [ 'CARTESIAN '] { x y z D }*
areaSpec    :=  rectSpec | circleSpec | hullSpec | convexSpec
regionSpec  := 'REGION ' {areaSpec}* | areaSpec
```

To give examples of region specifications:

**CIRCLE**  A point specification and a 1.75 nautical mile (arc minute) radius.
```
'CIRCLE LATLON 39.3 -76.61 100'
'CIRCLE CARTESIAN 0.1792 -0.7528 0.6334 100'
```
**RECT**  Two corner points defining the minimum and maximum of the lat, lon. The longitude coordinates are interpreted in the wrap-around sense, i.e., $lon_{min}$=358.0 and $lon_{max}$=2.0, is a 4 degree wide range. The latitudes must be between the North and South Pole. The rectangle edges are constant latitude and longitude lines, rather than the great-circle edges of CHULL and CONVEX.
```
'RECT LATLON 37 -109.55  41 -102.05'
```
**CHULL**  Three or more point specifications define a spherical convex hull with edges of the convex hull connecting adjacent points by great circles. The points must be in a single hemisphere, otherwise an error is returned. The order of the points is irrelevant.
```
'CHULL LATLON 37 -109.55 41 -109.55 41 -102.051 37 -102.05'
```
**CONVEX**  Any number (including zero) of constraints in the form of a Cartesian vector (*x,y,z*) and a fraction of the unit length of the vector.
```
'CONVEX    -0.17886 -0.63204 -0.75401 0.00000
           -0.97797  0.20865 -0.00015 0.00000
            0.16409  0.57987  0.79801 0.00000
            0.94235 -0.33463  0.00000 0.00000'
```
**REGION**  A region is the union of zero or more circle, rect, chull, and convex areas.
```
'REGION CONVEX 0.7 0.7 0.0 -0.5 CIRCLE LATLON 18.2 -22.4 1.75'
```

Any of these region descriptions can be fed to the `fHtmCoverRegion()` routine that returns a trixel table describing a set of trixels (triangular areas) covering that region. The simpler code for the Colorado query is:

```sql
select S.*
from   ( select ObjID
           from fHtmCoverRegion('RECT LATLON 37 -109.55  41 -102.05')
                loop join SpatialIndex
             on HtmID between HtmIdStart and HtmIdEnd
            and lat between 37 and 41
            and lon between -109.05 and -102.048
            and type = 'S') as G
  join Station S on G.objID = S.StationNumber
OPTION( FORCE ORDER)
```

This unusual query format is required to tell the optimizer exactly the order in which to perform the join (to make the "force order" option work correctly). It is difficult to modify the optimizer in this way, but until table-valued functions have statistics, they are estimated to be very expensive. You have to force them into the inner loop join.

The query returns 1030 stream gauges and has 1,365 candidates from the cover, so there are 25 percent false positives. Note that the rectangle cover is better than the circular cover, which had 61% false positives. There is polygon syntax for non-rectangular states, but this article is about table valued functions, not about the HTM algorithms. You can see the HTM code in the project, and also in the documentation for the project.

A similar query can be cast as a convex hull as:

```sql
select S.*
from   ( select ObjID
           from fHtmCoverRegion(
              'CHULL LATLON 37 -109.55 41 -109.55 41 -102.05 37 -102.05')
                loop join SpatialIndex
             on HtmID between HtmIdStart and HtmIdEnd
            and lat between 37 and 41
            and lon between -109.05 and -102.048
            and type = 'S') as G
  join Station S on G.objID = S.StationNumber
OPTION( FORCE ORDER)
```

The query returns 1030 stream gauges and has 1,193 candidates from the cover, so there are 14 percent false positives. The convex hull cover is even better than the equivalent rectangular cover in this case.

## 4. Advanced topics – complex regions.

The previous examples gave the syntax for regions and a discussion of point-near-point and point-in-rectangle searches. Regions can get quite complex. They are Boolean combinations of convex areas. We do not have the space here to explain regions in detail, but the HTM library in the accompanying project has the logic to do Boolean combinations of regions, simplify regions, compute region corner points, compute region areas, and has many other features. Those ideas are described in [Fekete], [Gray], and [Szalay].

To give a hint of these ideas, consider the state of Utah. Its boundaries are approximately defined by the union of two rectangles:
```sql
declare @utahRegion varchar(max)
set @utahRegion = 'region '
    + 'rect latlon 37 -114.0475  41 -109.0475 '  -- main part
    + 'rect latlon 41 -114.0475  42 -111.01    '  -- Ogden & Salt Lake.
```

Now we can find all stream gauges in Utah with the query:
```sql
select S.*
from  (
   select ObjID
   from fHtmCoverRegion(@utahRegion)
      loop join SpatialIndex
      on HtmID between HtmIdStart and HtmIdEnd
      and (((    lat between 37           and       41)  -- careful test
             and (lon between -114.0475 and -109.04))  -- are we inside
         or ((   lat between 41           and       42)  -- one of the two
             and (lon between -114.0475 and -111.01))   -- boxes?
            )
      and type = 'S' ) as G
  join Station S on G.objID = S.StationNumber
OPTION( FORCE ORDER)
```

The cover returns 38 trixels. The join returns 775 stations. The careful test finds 670 stations in Utah, and two Wyoming stations that are right on the border (14 percent false positives).

Most states require much more complex regions. For example, a region string to approximate California is:
```sql
declare @californiaRegion varchar(max)
set @californiaRegion = 'region '
              + 'rect   latlon 39     -125 '     -- nortwest corner
                         + '42     -120 '        -- center of Lake Tahoe
              + 'chull latlon 39     -124 '      -- Pt. Arena
                         + '39     -120 '        -- Lake  tahoe.
                         + '35     -114.6 '      -- start Colorado River
                         + '34.3   -114.1 '      -- Lake Havasu
                         + '32.74  -114.5 '      -- Yuma
                         + '32.53  -117.1 '      -- San Diego
                         + '33.2   -119.5 '      -- San Nicholas Is
                         + '34     -120.5 '      -- San Miguel Is.
                         + '34.57  -120.65 '     -- Pt. Arguelo
                         + '36.3   -121.9 '      -- Pt. Sur
                         + '36.6   -122.0 '      -- Monterey
                         + '38     -123.03 '     -- Pt. Rayes
select stationNumber
from fHtmCoverRegion(@californiaRegion)
   loop join SpatialIndex
   on HtmID between HtmIdStart and HtmIdEnd
   /* and <careful test> */
   and type = 'S'
 join Station S on  objID = S.StationNumber
OPTION( FORCE ORDER)
```

The cover returns 108 trixels, which cover 2,132 stations. Of these, 1,928 are inside California, so the false positives are about 5 percent -- but the careful test is nontrivial.

That same query, done for places rather than stations, with the careful test included, looks like this:

```sql
select *
from Place
where HtmID in
      ( select distinct SI.objID
       from fHtmCoverRegion(@californiaRegion)
             loop join SpatialIndex SI
                 on SI.HtmID between HtmIdStart and HtmIdEnd
                    and SI.type = 'P'
             join place P on SI.objID = P.HtmID
         cross join fHtmRegionToTable(@californiaRegion) Poly
        group by SI.objID, Poly.convexID
     having min(SI.x*Poly.x + SI.y*Poly.y + SI.z*Poly.z – Poly.d) >= 0
      )
OPTION( FORCE ORDER)
```

This uses the convex-halfspace representation of California and the techniques described in [Gray] to quickly test if a point is inside the California convex hull. It returns 885 places, seven of which are on the Arizona border with California (the polygon approximates California). It runs in 0.249 seconds on a 1GHz processor. If you leave off the "`OPTION( FORCE ORDER)`" clause it runs slower, taking 247 seconds.

Because this is such a common requirement, and because the code is so tricky, we added a procedure `fHtmRegionObjects(Region,Type)` that returns object IDs from SpatialIndex. This procedure encapsulates the tricky code above, so the two California queries become:

```sql
select *                  -- Get all the California River Stations
from Station
where stationNumber in    -- that are inside the region
 (select ObjID
  from fHtmRegionObjects(@californiaRegion,'S'))

select *                  -- Get all the California Cities
from Place
where HtmID in            -- that are inside the region
 (select ObjID
  from fHtmRegionObjects(@californiaRegion,'P'))
```

The Colorado and Utah queries are also simplified by using this routine.

## 4. Summary

The HTM spatial indexing library presented here is interesting and useful in its own right. It is a convenient way to index data for point-in-polygon queries on the sphere. But, the library is also a good example of how SQL Server and other database systems can be extended by adding a class library that does substantial computation in a language like C#, C++, Visual Basic, or Java. The ability to implement powerful table-valued functions and scalar functions and integrate these queries and their persistent data into the database is a very powerful extension mechanism that starts to deliver on the promise of Object-Relational databases. This is just a first step. In the next decade, programming languages and database query languages are likely to get even better data integration. This will be a boon to application developers.

# Appendix: The Basic HTM Routines

This section describes the HTM routines. The companion document [Szalay] has a manual page for each routine, and the routines themselves are annotated to support Intellisense.

In what follows, lat and lon are in decimal degrees (southern and western latitudes are negative), and distances are in nautical miles (arc minutes.)

### HTM library version: fHtmVersion() returns versionString

The routine returns an nvarchar(max) string giving the HTM library version.
Example use:
```
print dbo.fHtmVersion()
```
Returns something like:
```
'C# HTM.DLL V.1.0.0 1 August  2005 '
```

### Generating HTM keys: fHtmLatLon (lat, lon) returns HtmID

The routine returns the 21-deep HTM ID of that LatLon point.
Example use:
```
update Place set HtmID = dbo.fHtmLatLon(Lat,Lon)
```
There are also `fHtmXyz()` and `fHtmEq()` functions for astronomers.

### LatLon to XYZ: fHtmLatLonToXyz (lat,lon) returns Point (x, y, z)

The routine returns the Cartesian coordinates of that Lat Lon point.
Example use (this is the identity function):
```
Select LatLon.lat, LatLon.lon-360
from fHtmLatLonToXyz(37.4,-122.4) as XYZ cross apply
     fHtmXyzToLatLon(XYZ.x, XYZ.y, XYZ.z) as LatLon
```
There is also an `fHtmEqToXyz()` functions for astronomers.

### XYZ to LatLon: fHtmXyzToLatLon (x,y,z) returns Point (lat, lon)

The routine returns the Cartesian coordinates of that Lat Lon point.
Example use (this is the identity function):
```
Select LatLon.lat, LatLon.lon-360
from fHtmLatLonToXyz(37.4,-122.4) as XYZ cross apply
     fHtmXyzToLatLon(XYZ.x, XYZ.y, XYZ.z) as LatLon
```
There is also an `fHtmXyzToEq()` functions for astronomers.

### Viewing HTM keys: fHtmToString (HtmID) returns HtmString

Given an HtmID, the routine returns a nvarchar(32) in the form [N|S]$t_1t_2t_3…t_n$ where each triangle number $t_i$ is in {0,1,2,3} describing the HTM trixel at that depth of the triangular mesh. .
Example use:
```
print 'SQL Server development is at: ' +
               dbo.fHtmToString(dbo.fHtmLatLon(47.646,-122.123))
```
which returns: `'N132130231002222332302'`.
There are also `fHtmXyz()` and `fHtmEq()` functions for astronomers.

### HTM trixel Centerpoint: fHtmToCenterpoint(HtmId) returns Point (x, y, z)

Returns the Cartesian center point of the HTM trixel specified by the HtmID.
Example use:
```sql
select * from fHtmToCenterPoint(dbo.fHtmLatLon(47.646,-122.123))
```

### HTM trixel corner points: fHtmToCornerpoints(HtmId) returns Point (x, y, z)

Returns the three Cartesian corner points of the HTM trixel specified by the HtmID.
Example use:
```sql
select * from fHtmToCornerPoints(dbo.fHtmLatLon(47.646,-122.123))
```

### Computing distances: fDistanceLatLon(lat1, lon1, lat2, lon2) returns distance

Computes the distance, in nautical miles (arc minutes) between two points.
Example use:
```sql
    declare @lat float, @lon float
    select @lat = lat, @lon = lon
    from  Place
    where PlaceName = 'Baltimore' and State = 'MD'
    select PlaceName,
           dbo.fDistanceLatLon(@lat,@lon, lat,  lon) as distance
    from Place
```
There are also `fDistanceXyz()` and `fDistanceEq()` functions for astronomers.

The following routines return a table which serves as a spatial index. The returned spatial index table has the data definition:
```sql
SpatialIndexTable table (
    HtmID    bigint   not null ,  -- HTM spatial key (based on lat/lon)
    Lat      float    not null ,  -- Latitude in Decimal
    Lon      float    not null ,  -- Longitude in Decimal
    x        float    not null ,  -- Cartesian coordinates,
    y        float    not null ,  -- derived from lat-lon
    z        float    not null ,  --,
    Type     char(1)  not null ,  -- place (P) or gauge (G)
    ObjID    bigint   not null ,  -- object ID in table
    distance float    not null ,  -- distance in arc minutes to object
    primary key (HtmID, ObjID) )
```

### Finding nearby objects: fHtmNearbyLatLon(type, lat, lon, radius) returns SpatialIndexTable

Returns a list of objects within the radius distance of the given type and their distance from the given point. The list is sorted by nearest object.
Example use:
```sql
  select distance, Place.*
  from fHtmNearbyLatLon('P', 39.3, -76.6, 10) I join Place
  on I.objID = Place.HtmID
  order by distance
```
There are also `fHtmGetNearbyEq ()` and `fHtmGetNearbyXYZ()` functions for astronomers.

**Finding the nearest object:  fHtmNearestLatLon(type, lat, lon) returns SpatialIndexTable**

Returns a list containing the nearest object of the given type to that point.
Example use:
```sql
select distance, Place.*
from fHtmNearestLatLon('P', 39.3, -76.6) I join Place
on I.objID = Place.HtmID
```
There are also `fHtmGetNearestEq ()` and `fHtmGetNearestXYZ()` functions for astronomers.

The following routines return a table describing the HtmIdStart and HtmIdEnd of a set of trixels (HTM triangles) covering the area of interest.  The table definition is:
```sql
TrixelTable table (
          HtmIdStart   bigint not null ,  -- min HtmID in trixel
          HtmIdEnd     bigint not null    -- max HtmID in trixel
            )
```

**Circular region HTM cover:  fHtmCoverCircleLatLon(lat, lon, radius) returns trixelTable**

Returns a trixel table covering the designated circle.
Example use:
```sql
declare @answer nvarchar(max)
declare @lat float, @lon float
select @lat = lat, @lon = lon
from Place
where Place.PlaceName = 'Baltimore'
  and State = 'MD'
set @answer = ' using fHtmCoverCircleLatLon() it finds:     '
select @answer = @answer
        + cast(P.placeName as varchar(max)) + ', '
        + str( dbo.fDistanceLatLon(@lat,@lon, I.lat, I.lon) ,4,2)
        + '  arcmintes distant.'
from SpatialIndex I join fHtmCoverCircleLatLon(@lat, @lon, 5)
   On HtmID between HtmIdStart and HtmIdEnd  -- coarse test
   and type = 'P'                             -- it is a place
   and dbo.fDistanceLatLon(@lat,@lon, lat, lon)
                       between 0.1 and 5   -- careful test
   join Place P on I.objID = P.HtmID
    order by dbo.fDistanceLatLon(@lat,@lon, I.lat, I.lon) asc
print 'The city within 5 arcminutes of Baltimore is: '
    + 'Lansdowne-Baltimore Highlands, 4.37 arcminutes away'
```
There are also `fHtmCoverCircleEq()` for astronomers.

**General region specification to HTM cover: fHtmCoverRegion(region) returns trixelTable**

Returns a trixel table covering the designated region (regions are described earlier in this topic).
```sql
select S.*
from  ( select ObjID
        from fHtmCoverRegion('RECT LATLON 37 -109.55  41 -102.05')
              loop join SpatialIndex
            on HtmID between HtmIdStart and HtmIdEnd
           and lat between 37 and 41
           and lon between -109.05 and -102.048
           and type = 'S') as G
  join Station S on G.objID = S.StationNumber
```

```sql
OPTION( FORCE ORDER)
```

### General region simplification: fHtmRegionToNormalFormString(region) returns regionString

Returns a string of the form REGION {CONVEX {x y z d}* }* where redundant halfspaces have been removed from each convex; the convex has been simplified as described in [Fekete]

```sql
print dbo.fHtmToNormalForm('RECT LATLON 37 -109.55  41 -102.05')
```

The following routine returns a table describing the HtmIdStart and HtmIdEnd of a set of trixels (HTM triangles) covering the area of interest. The table definition is:

```sql
RegionTable ( convexID    bigint not null ,  -- ID of the convex, 0,1,…
             halfSpaceID bigint not null     -- ID of the halfspace
                                             -- within convex, 0,1,2,
             x           float  not null     -- Cartesian coordinates of
             y           float  not null     -- unit-normal-vector of
             z           float  not null     -- halfspace plane
             d           float  not null     -- displacement of halfspace
             )                               -- along unit vector [-1..1]
```

### Cast RegionString as Table: fHtmRegionToTable(region) returns RegionTable

Returns a table describing the region as a union of convexes, where each convex is the intersection of the x,y,z,d halfspaces. The convexes have been simplified as described in [Fekete]. Section 4 of this article describes the use of this function.

```sql
select *
from dbo.fHtmToNormalForm('RECT LATLON 37 -109.55  41 -102.05')
```

### Find Points Inside a Region: fHtmRegionObjects(region, type) returns ObjectTable

Returns a table containing the objectIDs of objects in SpatialIndex that have the designated type and are inside the region.

```sql
select *                   -- find Colorado places.
from Places join
where HtmID in
 select objID
 from dbo. fHtmRegionObjects('RECT LATLON 37 -109.55  41 -102.05','P')
```

### General region diagnostic: fHtmRegionError(region ) returns message

Returns "OK" if region definition is valid; otherwise, returns a diagnostic saying what is wrong with the region definition followed by a syntax definition of regions.

```sql
print dbo.fHtmRegionError ('RECT LATLON 37 -109.55  41 -102.05')
```

# SQL SERVER 2005
# HTM Interface Release 4
## Alex Szalay, Gyorgy Fekete, Jim Gray, August 2005

This document describes the SQL Server 2005 interfaces to the C# HTM functions. It also explains how to install and modify those procedures.  A tutorial on the Hierarchical Triangular Mesh (HTM) is at http://www.sdss.jhu.edu/htm/.

**Table of Contents**

# HTM Concepts Reviewed

The Hierarchical Triangular Mesh is a multilevel, recursive decomposition of the sphere. At the top, depth 1, there are eight spherical triangles, four each for the Northern and Southern hemispheres. Four of these triangles share a vertex at the pole. The sides opposite the pole form the equator. You can imagine these by orienting a regular octahedron so that two of its vertices are at the poles, and the other four are equally spaced on the equator. The spherical polygons are the projection of the edges of the octahedron onto the circumscribing sphere. There are eight unique integers that represent these triangles.

Triangles in the mesh scheme are called *trixels*. Each trixel can be split into four smaller trixels by introducing new vertices at the midpoints of each side, and adding a great circle arc segment to connect the new vertices with the existing one. Trixel division repeats recursively and indefinitely to produce smaller and smaller trixels. Each trixel has a *level number* that corresponds to the number of times an original (octant) triangle had to be split. Points in this decomposition are represented by a leading 1 bit and then the level 0 trixel number [0..7] and then the successive sub-trixel numbers [0..3]. This gives each trixel and its center-point a unique 64 bit identifier, called an *HTM ID* (HtmID) that represents a particular *trixel* in the HTM hierarchy. The smallest valid HtmID is 8 – being the level 0 HtmID of the triangle 0. HtmIDs are numbered from level 0. The term *depth* tells how many levels are involved: *depth = level+1*.

Though the division process can continue indefinitely, the 64-bit representation runs out of bits at depth 31. Depth 25 is good enough for most applications—about 0.6 meter on the surface of the Earth or 0.02 arc seconds. The code here defaults to depth 21 (0.3 arc seconds). Note that this numbering scheme is not a complete cover on the positive integers, and not all bit patterns form valid HtmID numbers.

Some of the functions described here return a region, or area, on the sphere. The return value of these functions are referred to as *trixel tables,* which are tables of HtmID ranges for trixels that overlap or cover the region. The structure of these tables is quite simple: they are rows of two 64-bit numbers (BIGINTs) [Htm_start, Htm_stop] that are the starting and ending values of HtmIDs of trixels in the range. The HtmIDs in this context are always for trixels at depth 21.

The library supports three coordinate systems:
1. *LatLon* is the Greenwich Meridian spherical coordinate system of latitude and longitude *(lat, lon)* used by geographers.
2. *J2000* or *Equatorial* is the celestial right ascension and declination *(ra, dec)* spherical coordinate system used by astronomers (the vector pointing at the center of the Milky Way defines the intersection of the J2000 prime meridian with the J2000 equator).
3. *Cartesian* is a the unit vector representation of a sphere; a point on the sphere (either LatLon or J2000) has a corresponding unit vector. The North Pole is (0,0,1) and the prime meridian intersection with the equator is (1,0,0).

This library is installed in the sample Spatial database, which uses the LatLon coordinates. Many of the examples are geared to that. The astronomy examples are drawn from tables at http://skyserver.sdss.org/.

# Find Version of Installed HTM Code

## fHtmVersion()
Returns a string describing version of HTM code. A typical description has a version and timestamp:
"`C# HTM.DLL V.1.0.0; 2005 July 30`".
Returns:
version varchar(max)      version string of installed code
Example use:
```
declare @version varchar(max)
select @version = dbo.fHtmVersion()
print 'Installed version is: ' + @version
```
Produces:
`Installed version is: C# HTM.DLL V.1.0.0; 2005 July 30`

# Geometric Conversion Functions

## fHtmXyzToLatLon(x, y, z)
Given the Cartesian coordinates (x, y, z), returns a table containing the corresponding LatLon coordinates.
Parameters:
```
x:  float not null      x
y:  float not null      y
z:  float not null      z
```

Returns:
VertexTable(lat float, lon float)    The lat/lon equivalent of the given point.
Example use:
```
Select * from dbo.fHtmXyzToLatLon(1.0, 0.0, 0.0)
```

Errors: Empty table is returned if (x, y, z) is too close to (0, 0, 0) (within 1e9 of zero).

## fHtmXyzToEq(x, y, z)
Given the Cartesian coordinates (x, y, z), returns a table containing the corresponding J2000 (ra, dec) coordinates.
Parameters:
```
x:  float not null      x
y:  float not null      y
z:  float not null      z
```

Returns:
VertexTable(ra float, dec float)    The RA/DEC equivalent of the given point.
Example use:
```
Select * from dbo.fHtmXyzToEq(1.0, 0.0, 0.0)
```

Errors: Empty table is returned, if (x, y, z) is too close to (0, 0, 0) (within 1e9 of zero).

## fHtmLatLonToXyz(lat, lon)
Given a LatLon point, returns a table containing the corresponding Cartesian coordinates (x, y, z).
Parameters:
```
lat:  float not null      latitude
```

```
    lon:   float not null        longitude
```

Returns:
```
VertexTable(x float, y float, z float)
```
Example use:
```
    Select * from dbo.fHtmLatLonToXyz(0.0, 0.0)
```

Errors: None. Extreme latitude values are truncated to [-90 … 90]

### fHtmEqToXyz(ra, dec)

Given a J2000 (equatorial) point returns a table containing the corresponding Cartesian coordinate (x, y, z).
Parameters:
```
    ra:   float not null        right ascension
    dec:  float not null        declination
```

Returns:
```
VertexTable(x float, y float, z float)
```
Example use:
```
    Select * from dbo.fHtmEqToXyz(0.0, 0.0)
```

Errors: None. Extreme declination values are truncated to [-90 … 90]

## Compute HtmID for a point

### fHtmLatLon (lat, lon)

Given a LatLon point, returns the 21-deep HtmID of that point on the earth.
Parameters:
```
    lat:   float not null        latitude
                                 It is converted to the range [-90 ... 90]
    lon:   float not null        longitude
```
Returns:
```
    htmID bigint not null        the 21-deep HtmID of that (ra,dec) point.
```
Example use:
```
    Declare @htmID bigint
    Select @htmID =dbo.fHtmLatLon(lat,lon)
    From Place
    where placeName = 'Baltimore' and state = 'MD'
```
Errors: None.

### fHtmEq(ra, dec)

Given a J2000 (equatorial) point, returns the 21-deep HtmID of that point on the celestial sphere.
Parameters:
```
    Ra:   float not null         right ascension in degrees.
                                 It is converted to a [0 … 360) range
    Dec:  float not null         declination in degrees.
                                 It is converted to the range [-90 ... 90]
```
Returns:
```
    htmID bigint not null        the 21-deep HtmID of that (ra,dec) point.
```
Example use:
```
    Declare @htmID bigint
    Select @htmID =dbo.fHtmEq(ra,dec)
    From Stars
    Where id = 42
```

Errors: None.

## fHtmXyz(x,y,z)

Given a Cartesian point, returns the 21-deep HtmID of that point on the sphere.
Parameters:
    `x:   float not null`                    vector to galaxy center or prime meridian intersection with equator
    `y:   float not null`                    vector normal to $x$ in galactic plane or normal to prime meridian
    `z:   float not null`                    vector normal to galactic plane or equator
    xyz will be normalized to 1.   ( (0, 0, 0) is converted to RA = 0, DEC = 0, i.e. (1, 0, 0)).
Returns:
    `htmID bigint not null`         the 21-deep HtmID of that (x,y,z) point.
Example use:
```
    Declare @htmID bigint
    Select @htmID = dbo.fHtmXyz(1,0,0)
```
Errors:  None.

# HTM Triangle Functions

## fHtmToCenterPoint (HtmID)

Given an HtmID, return its Cartesian x,y,z centerpoint as a vertex table.
Parameters:
    `HtmID: bigint not null`    HtmID of a triangle.
Returns:
    `VertexTable(x float, y float, z float)`.
Example use:
        `Select x,y,z from fHtmToCenterPoint(dbo.fHtmLatLon(38,115))`
Errors: None.

## fHtmToCornerPoints (HtmID)

Given an HtmID, returns the three Cartesian x,y,z corner points of the triangle as a vertex table. If the HtmID has less shallow depth, this will be a large triangle. For example, HtmID = 8 returns the corner points of the entire octant.
Parameters:
    `HtmID: bigint not null`    HtmID of a triangle.
Returns:
    `VertexTable(x float, y float, z float)`.
Example use:
        `Select x,y,z from fHtmToCornerPoints(dbo.fHtmLatLon(38,115))`
Errors: None.

# Distances Between Points

## fDistanceLatLon( lat1, lon1, lat2, lon2)

Given two LatLon points, fDistanceLatLon() returns distance between them in arc minutes (nautical miles).
Parameters:
    `lat1:  float not null`    latitude in degrees truncated to [-90, 90].
    `lon1:  float not null`    longitude in degrees truncated to [0,360).
    `lat2:  float not null`    latitude in degrees truncated to [-90, 90].
    `lon2:  float not null`    longitude in degrees truncated to [0,360).
Returns:
    `Float not null`    distance in arc minutes
Example use:
        `If (60 != dbo.fDistanceLatLon(0, 0, 1, 0) ) print 'error'`
Errors: None.

## fDistanceEq( ra1, dec1, ra2, dec2)

Given two J2000 (equatorial) points, fDistanceEq() returns distance between them in arc minutes.
Parameters:
    `Ra1:  float not null`    right ascension in degrees truncated to [0,360].
    `Dec1: float not null`    declination in degrees truncated to [-90, 90].
    `Ra2:  float not null`    right ascension in degrees truncated to [0,360].
    `Dec2: float not null`    declination in degrees truncated to [-90, 90].  .
Returns:
    `Float not null`    distance in arc minutes
Example use:
        `If (60 != dbo.fDistanceEq(0, 0, 1, 0) ) print 'error'`

Errors: None.

## fDistanceXyz( x1,y1,z1, x2,y2,z2)

Given two Cartesian points, fHtmXyz returns distance between them in arc minutes.
Parameters:
    x1, x2:   float not null      vector to galaxy center or prime meridian intersection with equator.
    y1, y2:   float not null      vector normal to *x* in galactic plane or prime meridian
    z1, z2:   float not null      vector normal to galactic plane or to north pole.
    (x, y, z) will be normalized. (0,0,0) will be converted to (0, 0, 1)

Returns:
    Float not null                      distance in arc minutes

Example use:
```
    If (5400 != dbo.fDistanceXyz(0, 1, 0, 0, 0, 1) ) print 'error'
```
Errors:  None.

# REGIONS

A region is an area of interest on the celestial sphere. You can specify a region as a polygon, a convex hull of a polygon, a rectangle, or a circle. Inside the kernel of the HTM engine, all regions are represented as a union of *convexes*, which are, in turn, intersections of halfspaces. For more information, see the article (Htm.doc) in the Geospatial project.

Syntactically, a region is a list of convexes. Furthermore, a convex is a list of halfspaces, and a halfspace is a 4-tuple {x, y, z, D}.

## Region Specifications

This is the syntax for region (cover) specifications:

```
circleSpec   := 'CIRCLE J2000'          ra  dec  rad
             |  'CIRCLE LATLON'         lat lon  rad
             |  'CIRCLE [CARTESIAN ]'   x y z    rad
rectSpec     := 'RECT J2000'            {ra  dec}2
             |  'RECT LATLON'           {lat lon}2
             |  'RECT [CARTESIAN ]'     {x y z  }2
polySpec     := 'POLY J2000'            {ra  dec}3+
             |  'POLY LATLON'           {lat lon}3+
             |  'POLY [CARTESIAN ]'     {x y z  }3+
hullSpec     := 'CHULL J2000'           {ra  dec}3+
             |  'CHULL LATLON'          {lat lon}3+
             |  'CHULL [CARTESIAN ]'    {x y z  }3+
convexSpec   := 'CONVEX J2000'          {ra  dec D}*
             |  'CONVEX LATLON'         {lat lon D}*
             |  'CONVEX [CARTESIAN ]'   {x y z   D}*
areaSpec     := circleSpec
             |  rectSpec
             |  polySpec
             |  hullSpec
             |  convexSpec
regionSpec   := 'REGION' {areaSpec}*
             |  areaSpec
```

## Examples of Region Specifications

**REGION** A number of convexes (including zero)
```
REGION CONVEX 1 0 0 0.7 0 1 0 0.7
REGION CONVEX J2000 0 0 0.99    5 3 0.99
REGION CONVEX J2000 0 0 0.99 CONVEX J2000 5 3 0.99
REGION CONVEX LATLON 90 0 0
REGION
```

**CONVEX** Any number of (including zero) constraints
```
CONVEX CARTESIAN 0.7 0.7 0.0 -0.5 0.7 -0.7 0.0 -0.5
CONVEX
```

**CIRCLE** A Point specification, like `J2000 ra,dec`, and an arc minutes radius. Angles are in degrees. Represented as a `CONVEX` consisting of a single constraint.
```
CIRCLE J2000 182.25 -22.432 1.75
CIRCLE CARTESIAN 0.7 0.0 0.7 1.75
```

**RECT**   Followed by two angular point specs, defining the minimum and maximum of the ra,dec. The $lat_{min}$ must be smaller than $lat_{max}$. In a similar case for the longitudes, they are interpreted in the wrap-around sense, i.e., $ra_{min}$=358.0 and $ra_{max}$=2.0, means a four-degree wide range.
```
RECT J2000 182.25 -1.432 184.75 1.44
```

**POLY**   Followed by an optional single coordinate specification and a number of corresponding point specifications (two or three numbers each). The spherical polygon will be created by connecting the points by great circle segments. Because it is restricted to a convex polygon, the order does not matter, but must be consistent. If there is a bowtie pattern in the points, or if the polygon is not convex, an error will result.
```
POLY J2000 -109.55 41 -102.05 41 -102.05 37 -109.55 37
```

**CHULL**   Followed by an optional single coordinate specification, and a number of corresponding point specifications (two or three numbers each). The spherical convex hull will be created by connecting the adjacent points by great circles. At least three points are needed. The points should all be within a single hemisphere, otherwise an error is returned. The order of the points is irrelevant.
```
CHULL J2000 180. -1. 190. -2. 185. 3. 182. 4. 185. 5.
```

## fHtmRegionToNormalFormString (regionSpec)

Given a string describing a region, fHtmRegionToNormalFormString () returns the normalized representation of that region as a union of non-empty convex hulls, with redundant constraints (halfspaces) discarded from each convex.

Parameters:
    regionSpec: `nvarchar(max) not null`  see syntax for region specifications above.

Returns:
    `nvarchar(max) not null`  returns the normalized region spec of the form
                                            REGION {CONVEX {x y z d}* }*
                                              (or null if error).

Example use:
```sql
Declare @regionSpec nvarchar(max)
Select @regionSpec = dbo.fHtmToNormalForm('CIRCLE J2000 195 0 1')
```

Errors:
    regionSpec syntax error, returns empty string, see fHtmRegionError()

## fHtmRegionToTable (regionSpec)

Given a string describing a region, fHtmRegionToTable() returns the tabular representation of the region as a union of non-empty convex hulls, with redundant constraints (halfspaces) discarded from each convex. The tabular representation has the schema described below:

Parameters:
    regionSpec: `nvarchar(max) not null`  see syntax for region specifications above.

Returns:
```
RegionTable (
        convexID    bigint not null ,   -- ID of the convex, 0,1,…
        halfSpaceID bigint not null     -- ID of the halfspace
                                        -- within convex, 0,1,2,
        x           float  not null     -- Cartesian coordinates of
        y           float  not null     -- unit-normal-vector of
        z           float  not null     -- halfspace plane
        d           float  not null     -- displacement of halfspace
                                        -- along unit vector [-1..1]
           )
```
    Or empty table if error.

Example use:

```sql
select *
from fHtmToNormalForm('CIRCLE J2000 195 0 1')
```

Errors:
    regionSpec syntax error, returns empty table., see fHtmRegionError()

## fHtmRegionObjects (regionSpec, type)

This routine is particular to the SQL Server 2005 sample spatial database and library which has Place, Station, and SpatialIndex tables, and has these functions installed. Given a string describing a region and a type "P" for place or "S" for station, fHtmRegionObjects() returns the tabular list of all the SpatialIndex objects of that type that are inside that region:

Parameters:
    regionSpec: `nvarchar(max) not null`  see syntax for region specifications above.
    Type: char(1):  "P" for Places in the Place table, "S" for Stations in the Station table

Returns:
```
ObjectTable (
        objID bigint not null primary key ,  -- ID of the object,…
                -- if type is "S", it is the Station.stationNumber
                -- if type is "T", it is the Place.HtmID
```

Or empty table if error.

Example use:
```sql
select *                    -- find Colorado places.
from Place
where HtmID in
 (select objID
  from fHtmRegionObjects('RECT LATLON 37 -109.55  41 -102.05','P'))
```
Errors:
   regionSpec syntax error, returns empty table., see fHtmRegionError()

## fHtmRegionError(regionSpec)

Returns "OK" if a valid regionSpec, else returns syntax error message.

Parameters:
   regionSpec: nvarchar(max) not null  see syntax for region specifications above.
Returns:
   nvarchar(max) not null      diagnostic message
Example use:
```sql
Declare @diagnostic nvarchar(max)
Select @ diagnostic =
       dbo.fHtmRegionError('CIRCLE J2000 195 0')
```
Errors: None.

# HTM Covers: Compute HtmID Ranges for a Region

This suite of routines, given a region specification, returns a table of trixels. The trixels cover the specified region. The trixels are described by a start-stop HTM pair. All points within the trixel are between the start-stop of the 21-deep HTM pair; in fact, they are in the closed interval [Htm_Start, Htm_Stop].
The resulting table has the definition:
    TrixelTable(Htm_Start bigint, Htm_Stop bigint)
Simple regions can be described as standard geometric shapes (circle, rectangle) giving the parameters. But typically, regions are described by using the linear syntax described above. Because the enumeration of HtmIDs tends to form locally connected intervals, the interface unifies these contiguous triangles as one large trixel.

## fHtmCoverRegion (regionSpec)

Given a string describing a region, fHtmRegionCover () returns the trixel table covering that region.
Parameters:
    regionSpec: nvarchar(max) not null   see syntax for region specifications above.
Returns:
    TrixelTable(Htm_Start bigint, Htm_Stop bigint)
Example use:

```
Select * from fHtmCoverRegion('CIRCLE J2000 195 0 1')
```

Errors:
    In case of error, returns the empty table. Use fHtmRegionError(RegionString) to get diagnostic.

## fHtmCoverCircleLatLon(lat, lon, radiusArcMin)

Given a (latitude, longitude) point and a radius in arc minutes, fHtmCoverCircleLatLon () returns the trixel table covering that circle.

Parameters:
    `Lat:   float not null`      latitude in degrees.
                                          It is converted to the range [-90 ... 90]
    `Lon: float not null`      longitude in degrees.
                                          `radiusArcMin: float not null` circle's radius in arc minutes.
    Radius should be positive and less than 180 degrees, i.e., 10800 minutes of arc.

Returns:
    `TrixelTable(Htm_Start bigint, Htm_Stop bigint)`

Example use:
        `Select * from fHtmCoverCircleLatLon(195,0,1)`

Errors: None:

## fHtmCoverCircleEq(ra, dec, radiusArcMin)

Given a J2000 ra, dec point and a radius in arc minutes, fHtmCoverCircleEq () returns the trixel table covering that circle.

Parameters:
    `Ra:   float not null`      right ascension in degrees.
                                          It is converted to a [0 … 360)
    `Dec: float not null`      declination in degrees.
                                          It is converted to the range [-90 ... 90]
    `radiusArcMin: float not null` circle's radius in arcminutes.
    Radius should be positive and less than 180 degrees, i.e., 10800 minutes of arc.

Returns:
    `TrixelTable(Htm_Start bigint, Htm_Stop bigint)`

Example use:
        `Select * from fHtmCoverCircleEq(195,0,1)`

Errors: None:

## fHtmCoverCircleXyz(x, y, z, radiusArcMin)

Given a string describing a region, fHtmCoverCircleXyz() returns the trixel table covering that circle.

Parameters:
    `x:   float not null`           vector to galaxy center
    `y:   float not null`           vector normal to x in galactic plane
    `z:   float not null`           vector normal to galactic plane
    `radiusArcMin: float not null` circle's radius in arc minutes.
    xyz will be normalized to 1. (0,0,0) is converted to the North Pole: (0,0,1).
    Radius is range limited to [0…10800]

Returns:
    `TrixelTable(Htm_Start bigint, Htm_Stop bigint)`

Example use:
        `Select * from fHtmCoverCircleXyz(1,0,0,1)`

Errors: None:

## Installing the HTM Code

1. Install the SQL Server 2005 samples by following the instructions in the **Installing Samples** topic in SQL Server Books Online. By default the sample will be installed in *drive*:\Program Files\Microsoft SQL Server\90\Samples\Engine\Programmability\CLR\Spatial\, where *drive* is the system drive.

2. Compile the provided solution by using Visual Studio 2005 or the Microsoft .NET Framwork SDK 2.0 using a command similar to the following in a .NET Framework SDK command prompt:

   ```
   msbuild /property:configuration=debug CS\Spatial.sln
   ```

3. Attach the Spatial database in the `data` directory by using SQL Server Management Studio or by executing the `Scripts\AttachSpatialDatabase.bat` command file in a command prompt window if you have not done so already.

4. Execute the spHtmCsharp.sql script in SQL Server Management Studio, or by executing a command similar to the following in a command prompt window:

   ```
   sqlcmd –S "(local)" –d Spatial -E -i "C:\Program Files\Microsoft SQL Server\90\Samples\Engine\Programmability\CLR\Spatial\Scripts\spHtmCsharp.sql"
   ```

This script requires SQL Server 2005 be the default database server on the local system. This command is part of the contents of `Scripts\BuildSpatialDatabase.bat`. It does not work with SQL Server 2000 or earlier versions. The `spHtmCsharp.sql` script enables the common language runtime (CLR), drops any existing HTM assembly and replaces it with the current assembly (`CS\Spatial\bin\debug\Spatial.dll`), and then defines all the HTM functions in that assembly.

## Compiling, Modifying, and Debugging the HTM Code

The C:\HtmCsharp\htm.sln is a C# database project. You should be able to use the "deploy" and the "debug" features and breakpoints will just work.

# Table of HTM Depths And Approximate Areas

The library here defaults to 21-deep HTM keys (the first level divides the sphere into eight faces, and each subsequent level divides the spherical triangle into four sub-triangles.) The table below indicates that a 21-deep trixel is fairly small. The code can be modified to go 31-deep deep before the 64-bit representation runs out of bits, but the floating point representation and transcendental functions lose precision near level 25 .

| HTM depth | Area | | | | trixel | Objects/Trixel | |
|---|---|---|---|---|---|---|---|
| | degees$^2$ | minute$^2$ | area sec$^2$ | earth meters$^2$ | | SDSS | USGS |
| sphere | 41253 | 148,510,800 | 534,638,880,000 | 5.E+14 | | | |
| 0 | 5157 | 18,563,850 | 66,829,860,000 | 6E+13 | | 3E+8 | |
| 1 | 1289 | 4,640,963 | 16,707,465,000 | 2E+13 | | 8E+7 | |
| 2 | 322 | 1,160,241 | 4,176,866,250 | 4E+12 | | 2E+7 | |
| 3 | 81 | 290,060 | 1,044,216,563 | 1E+12 | | 5E+6 | |
| 4 | 20 | 72,515 | 261,054,141 | 2E+11 | | 1E+6 | 30,000 |
| 5 | 5 | 18,129 | 65,263,535 | 6E+10 | | 3E+5 | 7,500 |
| 6 | 1 | 4,532 | 16,315,884 | 2E+10 | 1 deg$^2$ | 73242 | 1,875 |
| 7 | 3E-1 | 1,133 | 4,078,971 | 4E+9 | | 18311 | 468 |
| 8 | 8E-2 | 283 | 1,019,743 | 1E+9 | | 4578 | 117 |
| 9 | 2E-2 | 71 | 254,936 | 2E+8 | | 1144 | 29 |
| 10 | 5E-3 | 18 | 63,734 | 6E+7 | | 286 | 7 |
| 11 | 1E-3 | 4 | 15,933 | 2E+7 | | 72 | 2 |
| 12 | 3E-4 | 1 | 3,983 | 4E+6 | 1 amin$^2$ | 18 | 0.5 |
| 13 | 8E-5 | 3E-1 | 996 | 943816 | | 4 | 0.1 |
| 14 | 2E-5 | 7E-2 | 249 | 235954 | | 1 | |
| 15 | 5E-6 | 2E-2 | 62 | 58989 | | 0.3 | |
| 16 | 1E-6 | 4E-3 | 16 | 14747 | | . | |
| 17 | 3E-7 | 1E-3 | 4 | 3687 | | | |
| 18 | 8E-8 | 3E-4 | 1 | 922 | | | |
| 19 | 2E-8 | 7E-5 | 2E-1 | 230 | 1 asec$^2$ | | |
| 20 | 5E-9 | 2E-5 | 6E-2 | 58 | 1 km$^2$ | | |
| 21 | 1E-9 | 4E-6 | 2E-2 | 14 | | | |
| 22 | 3E-10 | 1E-6 | 4E-3 | 4 | | | |
| 23 | 7E-11 | 3E-7 | 9E-4 | 1 | 1 m$^2$ | | |
| 24 | 2E-11 | 7E-8 | 2E-4 | 2E-1 | | | |
| 25 | 5E-12 | 2E-8 | 6E-5 | 6E-2 | | | |
| 26 | 1E-12 | 4E-9 | 1E-5 | 1E-2 | | | |

**Table 1:** Each HTM level subdivdies the sphere. For each level, this table shows the area in square degrees, arc minutes, arc seconds, and meters. The Trixel colum shows some characteristic sizes: the default 21-deep trixels is about .3 arc second$^2$. The USGS data has about ½ object per 12-deep trixel.